\documentclass[9.5pt, conference]{IEEEtran}
\IEEEoverridecommandlockouts
\usepackage{cite}
\usepackage{amsmath,amssymb,amsfonts}
\usepackage{algpseudocode}
\usepackage{algorithm}
\usepackage{graphicx}
\usepackage{textcomp}
\usepackage{xcolor}
\usepackage{tikz}
\usepackage{algpseudocode}
\usetikzlibrary{arrows.meta,positioning,fit,backgrounds}
\usepackage{enumitem}
\def\BibTeX{{\rm B\kern-.05em{\sc i\kern-.025em b}\kern-.08em
    T\kern-.1667em\lower.7ex\hbox{E}\kern-.125emX}}

\usepackage{booktabs}
\usepackage{array}
\usepackage{multirow}
\usepackage{url}
\usepackage{authblk}
\usepackage{caption}

\begin{document}

% \title{Be Careful Who You're Talking To:  The Importance of Speech Genre for End-of-Turn Detector Training}
% \title{Who You're Talking To Matters:  The Importance of Speech Genre for End-of-Turn Detector Training}
\title{The Importance of Speech Genre for End-of-Turn Detector Training}

\author{Rini Sharon$^{\star}$ \qquad Kadri Hacioglu$^{\star}$ 
\qquad Jeena Prakash$^{\star}$
\qquad Andreas Stolcke$^{\star}$}

\affil{$^{\star}$ Uniphore}
% \title{Speech Genre and Supervision Granularity in End-of-Turn Detector Training}
% \title{The Overlooked Pair: Speech Genre and Supervision Granularity in End-of-Turn Detection}

% \author{\IEEEauthorblockN{Anonymous Authors}
% \IEEEauthorblockA{\textit{Affiliation withheld for double-blind review}}}

% \author{\IEEEauthorblockN{Rini Sharon}
% \IEEEauthorblockA{\textit{Uniphore} \\
% India \\
% evanjilina@gmail.com}
% \and
% \IEEEauthorblockN{Andreas Stolcke}
% \IEEEauthorblockA{\textit{Uniphore} \\
% PA \\
% andreas@uni.com}
% \and
% \IEEEauthorblockN{Kadri Hacioglu}
% \IEEEauthorblockA{\textit{Uniphore} \\
% Colarado \\
% kadri@uni.com}
% }

\maketitle

\begin{abstract}
End-of-turn (EoT) detection has been studied mostly through features, models, and modalities, while the speech genre of the training data has drawn far less attention. We isolate this factor. Using one reproducible annotation pipeline, we build a matched pair of corpora, multi-turn human-human conversational and single-turn human-machine, and train the same streaming ConvLSTM EoT detector on each. Across four test sets the conversational model reaches up to +0.45 higher F1, driven by fewer false alarms on the mid-turn pauses and hesitations that are rare in single-turn corpora. This conversational-training effect is not specific to frame-level decoding. Under a matched complete-or-not test it also appears at the utterance level, across two encoders, but only on conversational test data. Testing only on single-turn speech, as most prior work does, therefore hides the benefit of conversational training data. Against open-source baselines, our 1.8M-parameter model attains the highest F1 on every test set.

\end{abstract}


\section{Introduction and Related Work}
\label{sec:intro_related}

Turn-taking is how speakers coordinate who holds the floor and when a turn has ended. In a spoken dialogue system the task comes down to deciding when the user has finished, so the system can respond. Responding too early interrupts the user, and waiting too long makes the exchange feel unnatural~\cite{raux2008,levinson2015}. A fixed silence threshold remains a widely used default~\cite{raux2008,schlangen2011}: it treats every pause the same and cannot tell a mid-turn hesitation from a true turn end~\cite{heldner2010,levinson2015}. To get past this, prior work has improved end-of-turn (EoT) detection along four directions. Three are well established, namely the features used as turn-taking cues, the modeling approach, and the signal modalities. The fourth and more recent direction is the composition of the training data. We review each in turn.

\noindent \textbf{Feature axis.} Early systems combined silence-based rules with prosodic cues~\cite{Ferrer2002SpeakerDone,ferrer03_icassp}. Later work showed that richer acoustic features such as pitch, energy, and duration help separate true completions from mid-turn pauses~\cite{skantze2017towards,roddy2018investigating}, and that higher-level cues beyond acoustics improve prediction further~\cite{aldeneh2018improving}.

\noindent \textbf{Modeling axis.} Approaches range from voice activity detection (VAD) with fixed pause thresholds~\cite{skantze2017towards,aldeneh2018improving}, to discrete classifiers placed at candidate boundaries~\cite{aldeneh2018improving}, to frame-level models that predict upcoming speech activity~\cite{skantze2017towards}. Voice activity projection (VAP) uses a transformer to predict joint voice-activity states in a self-supervised way~\cite{ekstedt2022vap}, and text-only models learn to predict turn-shift tokens from dialogue corpora~\cite{ekstedt2020turngpt}. More recent work moves toward semantic endpointing, where the decision depends on what the user said and not only on silence. One such design pairs a lightweight on-device detector with a stronger server model to tell a true turn end from a mid-turn hesitation~\cite{ok2025}. These detectors can run standalone or sit inside ASR and full-duplex spoken language systems~\cite{ekstedt2022vap,moshi2024,lin2025fdbench}.

\noindent \textbf{Modality axis.} Recent systems combine acoustic encoders with text and visual streams and outperform unimodal baselines on multi-party conversations~\cite{lee2024online,castillo2025survey}. Adding gaze, head pose, and facial action units gives further gains in face-to-face settings~\cite{visualcues2025,castillo2025survey}. Even so, a recent survey reports that 72\% of turn-taking studies do not benchmark against prior methods~\cite{castillo2025survey}, which makes it hard to say whether gains come from the model or from differences in data and evaluation.

\noindent \textbf{Data axis.} The training data itself shapes turn-taking behavior, yet it has drawn the least attention of the four axes. Many works train and evaluate on a single corpus with no cross-data comparison~\cite{castillo2025survey}. Multilingual VAP transfers poorly across languages when trained on a single language~\cite{inoue2024multilingual}, and work extending turn-taking to triadic and multi-party settings points to a similar sensitivity to the training corpus~\cite{triadicvap2025,lee2024online}. These difficulties reflect a broader shortage of diverse, real end-of-turn training data, which has prompted recent corpora that mix real and synthetic speech~\cite{ok2025}.

We focus on this last axis. The training data for EoT detection falls into broad speech genres that differ in the turn-taking behavior they contain.
\begin{enumerate}[nosep,leftmargin=*]
  \item \emph{Conversational multi-turn speech} is human-to-human dialogue, rich in hesitations, backchannels, overlapping speech, and mid-turn pauses~\cite{triadicvap2025,lee2024online,cieri2004}.
  \item \emph{Human-to-system dialogue} is a user addressing a system that prompts and then waits silently, so turns have cleaner boundaries~\cite{smartturn2025}.
  \item \emph{Command-and-control speech} is short single-turn queries with clean onsets and offsets~\cite{bastianelli2020}.
  \item \emph{Read or scripted speech} is prompted sentences or audiobooks with few turn-taking signatures~\cite{castillo2025survey}.
  \item \emph{Mixed real-plus-synthetic speech} is human recordings augmented with TTS corpora~\cite{smartturnv2blog,ultraVADreadme}.
\end{enumerate}

What separates these genres is how many within-turn events they expose, and that is exactly what a model can learn from. Conversational data are full of mid-turn pauses, filled pauses, and backchannels that should not trigger an EoT, and a model can learn to hold back at them from their prosodic and contextual cues~\cite{gravano2011turn,skantze2017towards}. The other genres carry far fewer such events. Human-to-system data lack the floor negotiation of human-human dialogue~\cite{raux2008flexible,castillo2025survey}, command-style data offer little within-turn material, and synthetic speech may not capture the full range of hesitation cues found in natural speech~\cite{castillo2025survey,edlund2017vocal}. The contrast is widened by evidence that speakers adjust their timing and prosody once they know they are addressing a machine~\cite{darves2002adaptation,oviatt2004adaptive,raux2008flexible}.

We therefore study two genres that contrast sharply on this within-turn axis: multi-turn human-human speech and single-turn human-machine speech. We pick these because large amounts of real speech can be reliably labelled for both, while the other combinations are either uninformative or hard to obtain. A single utterance with no second speaker offers no turn-taking to learn from, and real multi-turn human-machine dialogue with reliable end-of-turn labels remains scarce, so recent work has had to annotate it from scratch~\cite{mmf2f2025}. Restricting the comparison to these two lets us vary one factor, the density of within-turn events, while labelling both genres the same way. This matters because the two are rarely compared under matched conditions: studies that train on conversational data~\cite{ekstedt2022vap,skantze2017towards} seldom test against single-turn training, and systems built on single-turn or synthetic data~\cite{smartturn2025,ok2025} seldom test against conversational training. Genre therefore stays underexplored and entangled with differences in architecture, features, and evaluation, so a reported gain cannot be traced to the data itself.

% We therefore study two genres that contrast sharply on this within-turn axis, namely multi-turn human-human speech and single-turn human-machine speech. We pick these two because large amounts of real speech can be reliably labelled for both. The other combinations are either uninformative or hard to obtain. A single human utterance with no second speaker has no turn-taking to learn from, and real multi-turn human-machine dialogue with reliable end-of-turn labels is still scarce, so recent work has had to collect and annotate it from scratch~\cite{mmf2f2025}. Restricting the comparison to these two lets us vary one factor, the density of within-turn events, while labelling both genres the same way. This setup matters because the two genres are rarely compared under matched conditions. Studies that train on conversational data~\cite{ekstedt2022vap,skantze2017towards} seldom test against single-turn training, and systems built on single-turn or synthetic data~\cite{smartturn2025,ok2025} seldom test against conversational training. The effect of genre therefore stays underexplored and entangled with differences in architecture, features, and evaluation, so a reported gain cannot be traced to the data itself.

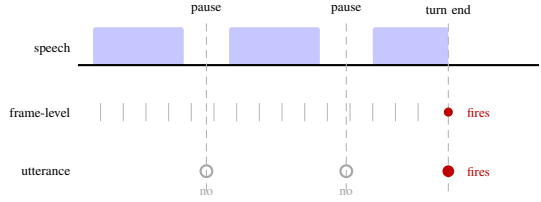
\begin{figure}[t]
\centering
\footnotesize
\begin{tikzpicture}[font=\scriptsize, >=Stealth]
% speech timeline: three chunks split by two mid-turn pauses; the turn ends after the last chunk
\foreach \a/\b in {0.2/1.4, 2.0/3.2, 3.9/4.9} {
  \fill[blue!22, rounded corners=1pt] (\a,0) rectangle (\b,0.5);
}
\draw[thick] (0,0) -- (6.1,0);
\node[left, font=\tiny] at (0,0.22) {speech};
% a dashed guide at each pause and at the turn end
\foreach \x in {1.7, 3.55, 4.9} {\draw[densely dashed, gray!60] (\x,0.55) -- (\x,-1.7);}
\node[font=\tiny] at (1.7,0.73) {pause};
\node[font=\tiny] at (3.55,0.73) {pause};
\node[font=\tiny] at (4.9,0.73) {turn end};
% frame-level: a decision at every frame, fires at the true end
\node[left, font=\tiny] at (0,-0.62) {frame-level};
\foreach \x in {0.3,0.6,...,4.8} {\draw[gray!55] (\x,-0.5) -- (\x,-0.74);}
\fill[red!75!black] (4.9,-0.62) circle (1.7pt);
\node[right, font=\tiny, red!75!black, anchor=west] at (5.05,-0.62) {fires};
% utterance-level: the classifier is queried once at each pause
\node[left, font=\tiny] at (0,-1.4) {utterance};
\draw[thick, gray!70] (1.7,-1.4) circle (2.2pt);
\node[font=\tiny, gray!70] at (1.7,-1.68) {no};
\draw[thick, gray!70] (3.55,-1.4) circle (2.2pt);
\node[font=\tiny, gray!70] at (3.55,-1.68) {no};
\fill[red!75!black] (4.9,-1.4) circle (2.2pt);
\node[right, font=\tiny, red!75!black, anchor=west] at (5.05,-1.4) {fires};
\end{tikzpicture}
% \caption{Frame-level decides at every frame. The utterance classifier is queried at each pause and fires at the first one judged complete.}
\caption{Frame-level vs.\ utterance-level decisions on a turn with two mid-turn pauses. The frame-level detector scores every frame, holds back through both pauses, and fires only at the turn end. The utterance-level classifier is queried only at a pause and commits at the first one it judges complete.}
\label{fig:concept}
\end{figure}

A second gap is about how the end-of-turn decision is produced. A frame-level detector emits a prediction at every frame, so it can hold back through a mid-turn pause and commit only at the turn end~\cite{skantze2017towards,roddy2018investigating,ekstedt2022vap}. An utterance-level detector instead acts only at a voice-activity pause, classifying the speech up to that pause as a complete or incomplete turn, so it makes a single decision per segment~\cite{aldeneh2018improving,smartturn2025}. Figure~\ref{fig:concept} contrasts the two. Because utterance-level systems are tested almost only on single-turn or clean speech~\cite{smartturnv3blog,bastianelli2020} and cross-genre comparison is rare~\cite{castillo2025survey}, whether conversational training helps them remains untested.

\noindent\textbf{Contributions.}
To our knowledge, this is the first study to isolate the \emph{speech genre of the training data} as a controlled variable for end-of-turn detection. Our three contributions measure this effect, bound its scope, and benchmark it against prior systems:

\begin{enumerate}
  \item \textbf{A controlled comparison of training genre.} Using one labelling pipeline, we build conversational and single-turn training sets matched in size and labelling and train the same detector on each, so training genre is the variable under test.
  \item \textbf{The scope of the effect.} We rerun this comparison at the utterance level, where most recent detectors operate~\cite{smartturn2025}, across conversational and single-turn test genres and two independently pretrained encoders, mapping where the effect holds and where it does not.
  \item \textbf{A benchmark with significance tests.} We compare our compact streaming detector against open-source VAD and learned baselines under one protocol with a one-sided tolerance window, a silence-matched utterance test, and significance tests.
\end{enumerate}

\section{Methodology}
\label{sec:experimental_design}

This section describes how we build the data, the model that consumes it, the training setup, the datasets we compare on, the baselines, and the metrics.

\subsection{Data Curation Pipeline}
\label{sec:training_data}
The conversational corpus comes from real human-to-human call center conversations in US English, spanning banking, insurance, retail, and telecommunications. Recordings are 8\,kHz stereo with one channel per speaker, the native rate of telephony, where the channel limits speech to roughly the 300 to 3400\,Hz band~\cite{itu-t-g712, Valin2000Bandwidth}. The cues that matter for end-of-turn detection sit well below this upper edge, since pitch is usually 80 to 400\,Hz and the energy and pause patterns we rely on are low-frequency as well, so 8\,kHz preserves them~\cite{rabiner2011speech}. For baselines that expect 16\,kHz input (Silero, SmartTurn, and Whisper), we upsample the audio before inference.

The curation pipeline runs in three stages, turn extraction, silence extension, and frame-level labelling.

\noindent\textbf{Turn extraction.} This stage splits each call into single speaker turns and marks the EoT at a turn's last frame, the point after which the other speaker takes over or the call ends. Algorithm~\ref{alg:fullcall} gives the steps.

\begin{algorithm}[t]
\caption{Full-Call EoT Annotation Pipeline}
\label{alg:fullcall}
\small
\begin{algorithmic}[1]
\State \textbf{Input.} Stereo calls \(\rightarrow\) per-channel mono WAV at 8kHz
\State \textbf{VAD + Smoothing.} Extract speech regions $R_i = [t_{start}, t_{end}]$ per channel, then smooth via gap-filling and noise-pruning (kernel size 3) \cite{sohn1999}
\State \textbf{Turn Merging.} Merge consecutive same-channel $R_i, R_{i+1}$ if gap $< 0.5$\,s and opponent silent (this reassembles VAD fragments into turns, and the EoT is placed in Step~5 only)
\State \textbf{Filtering.} Discard segments where $duration < 0.1\,$s (noise) OR ($duration < 0.2\,$s AND overlaps opponent, a backchannel)
\State \textbf{EoT Placement.} Mark EoT at $t_{end}$ if opponent speaks after turn end or turn is final
\State {\textbf{Output.} Full-call EoT annotations}
\end{algorithmic}
\end{algorithm}

\noindent\textbf{Silence extension.} This stage appends the silence that follows each turn. A sustained silence after the final word is a strong cue that the turn has ended~\cite{heldner2010,skantze2017towards}. The window extends up to 1.3\,s past the EoT and is VAD-validated to contain at most 10\% speech. If the next speaker or the same speaker resumes inside the window, the extension stops at that point. We keep a turn only when it has at least 0.3\,s of clean trailing silence, and we cap the window at the file boundary.

\noindent\textbf{Frame-level labelling.} This assigns binary labels at 10\,ms resolution. Frames before the EoT are labeled 0, and frames from the EoT through the appended silence are labeled 1.

% \noindent\textbf{Frame-level labelling.} This stage assigns binary labels at 10\,ms resolution. Frames before the EoT are labeled~0, and frames from the EoT through the appended silence are labeled~1.

\subsection{Datasets}
\noindent \textbf{Training datasets.} The central comparison rests on two 10K-utterance training sets, one conversational and one single-turn. \emph{Conv-10K} holds 10{,}000 segments drawn from multi-turn human-to-human call center conversations using the pipeline above. \emph{ST-10K} holds 10{,}000 complete single-turn human-to-system utterances, both real and synthetic, from the SmartTurn v3 training partition~\cite{smartturn2025}, drawn as a seeded, stratified random sample that is disjoint from the evaluation utterances. Because each utterance is a complete turn, its end is itself the turn end, so Stage~1 (turn extraction) does not apply. Frame-level labels then come from the same Stage~2 and Stage~3 as Conv-10K. The speech offset of each utterance is taken as its EoT, frames before it are labeled 0 and frames after it 1, and the same silence extension is applied. The labelling rule is thus identical for both genres. We refer to these as conversational and single-turn data throughout, and both models share the same architecture and hyperparameters.

\noindent \textbf{Evaluation datasets.} We evaluate on four test sets that span human-human conversation, human-machine dialogue, and single-command interaction, so each training regime is tested both in-domain and out-of-domain. All audio uses the same 8\,kHz feature extraction as training.

\begin{description}[leftmargin=*,nosep,style=unboxed]
\item[Internal 5K.] 5{,}000 speaker-separated segments from multi-turn customer service calls with human-verified EoT annotations. In-domain for Conv-10K.
\item[CASPER~\cite{hlavnickova2017casper}.] 2{,}500 turn segments from an independent open-source corpus of two-party English conversations with hesitations, backchannels, and mid-turn pauses, processed through the same full-call pipeline. It tests whether conversational training carries beyond the source domain.
\item[ST-Complete~\cite{smartturn2025}.] 2{,}500 samples (mean 9.12\,s) drawn from the SmartTurn v3 pool of complete turns (turns that have genuinely ended, i.e.\ with a true EoT). In-domain for ST-10K, which gives a symmetric cross-training comparison. Since SmartTurn ships only binary per-sample labels, frame-level labels come from VAD-based speech detection. SmartTurn reports above roughly 80\% accuracy on its own test sets~\cite{smartturnv3blog}, which include real and synthetic audio and both complete and incomplete turns. Our subset keeps only real complete turns, chosen to make the comparison with the conversational and open-source data fair, which accounts for the gap relative to the published SmartTurn numbers.
\item[SLURP~\cite{bastianelli2020}.] 2{,}500 short voice commands (mean 2.91\,s, test split, seed\,=\,42) with no multi-turn structure. It sits entirely outside the turn detection domain and tests whether an EoT detector carries over to command-and-control speech.
\end{description}

\subsection{Model Architecture}
\label{sec:architecture}
The model consumes the labelled frames one at a time. Each 10\,ms frame is an 85-dimensional vector that joins an 80-dim log-Mel filterbank with 5 prosodic features, namely normalized F0, F0 delta, a voiced flag, a speech flag, and cumulative silence duration.

A Conv1D (kernel size 33, 256 channels) gives 320\,ms of lookahead per frame. Its output passes through LayerNorm and a GELU~\cite{hendrycks2016gelu} activation, then into a 2-layer unidirectional LSTM (256 hidden units, dropout 0.25 between layers). A 3-layer MLP (256\,$\to$\,128\,$\to$\,64\,$\to$\,1, GELU activations, dropout 0.25) maps to a per-frame EoT logit, and a sigmoid converts this into a per-frame EoT probability. The model has 1.8M trainable parameters. Because the LSTM states carry over across chunks, the model can stream continuously with unbounded left context.

\subsection{Training Methodology}
\label{sec:training}
EoT frames are far rarer than non-EoT frames, so the loss is binary cross-entropy with a positive-class weight $w^{+} = {N_{\text{neg}}}/{N_{\text{pos}}}= 5.6$, equal to the ratio of non-EoT to EoT frames, which stops the model from trivially predicting the majority class. To match streaming inference, each sample is processed chunk by chunk with the LSTM hidden states carried forward, and the loss is masked to each sample's valid frames.

We optimize with Adam~\cite{kingma2015adam} at learning rate $3\times10^{-4}$ and weight decay $1\times10^{-4}$, with 2 epochs of linear warmup followed by cosine decay~\cite{loshchilov2016sgdr} to zero. Gradients are clipped at norm 5.0. Batch size is 8, with samples sorted by length to reduce padding. Training runs up to 30 epochs with early stopping on validation F1 (patience 7).

\subsection{Baselines}
\label{sec:baselines_desc}
We compare against three paradigms, a signal-based VAD, a neural VAD, and a learned end-of-turn classifier. The two VAD baselines decide from silence alone. Each fires an EoT once the silence after speech grows longer than a threshold, so neither can tell a mid-turn pause from a true turn end. We sweep this threshold from 0 to 1000\,ms in 50\,ms steps and report the best F1 per baseline per dataset. WebRTC VAD~\cite{webrtc} is the signal-based detector at aggressiveness level 3, and Silero VAD v5~\cite{silero2021} is the neural detector with its default threshold of 0.5.

The third baseline, SmartTurn v3~\cite{smartturn2025}, is a learned, utterance-level classifier rather than a silence rule. A Whisper-Tiny encoder~\cite{whisper2023} with a linear classifier on top (8M parameters in total) labels the speech up to a pause as a complete or incomplete turn, and Silero VAD supplies the pauses (Figure~\ref{fig:concept}). We also evaluate SmartTurn FT, the same model fine-tuned on our conversational data.

To check that any utterance-level effect is not tied to one model, we add two more utterance-level baselines, each a frozen pretrained encoder with only a small classifier trained on top, so the training genre of the classifier is the only variable. The two encoders, Whisper-Tiny~\cite{whisper2023} and wav2vec2-base~\cite{baevski2020}, are pretrained very differently, one on transcribed speech and one on untranscribed audio, so they encode a segment in independent ways. SmartTurn also builds on Whisper-Tiny, but it updates the whole encoder during training rather than keeping it fixed as we do. With the two SmartTurn baselines, this gives four utterance-level systems across two encoders.

% They check that our findings do not hang on a single architecture.

\subsection{Evaluation Metrics}
\label{sec:metrics}
A positive EoT decision fires when the sigmoid output stays above threshold $\tau$ for $d=8$ consecutive frames (80\,ms at 100\,Hz), and the predicted endpoint $\hat{t}$ is the last frame of this crossing. The tolerance window is one-sided: a detection is correct if it is no more than 50\,ms early ($-50\text{ms} \le (\hat{t} - t_{\text{gt}})$, with $t_{\text{gt}}$ the ground-truth boundary), and earlier triggers are false alarms, since a premature interruption costs more than a modest delay. We impose no late cutoff, as a late but correct endpoint is still correct, and report lateness through the detection latency $\delta = \hat{t} - t_{\text{gt}}$. Because a one-sided window could reward a model that simply waits, we read F1 together with $\delta$ and credit a high F1 only when latency stays near natural turn gaps.

In the frame-level setting, each system runs over the whole utterance and yields one endpoint $\hat{t}$: for our model, the first sustained threshold crossing; for the SmartTurn baselines, the first pause judged to be a turn end. However $\hat{t}$ is produced, it is scored against the ground-truth boundary with the same tolerance window, so all systems compare on equal terms. Two elements of this protocol are specific to our setting rather than standard practice. The tolerance window is one-sided, so an early fire is always a false alarm instead of being absorbed by a symmetric window, which is the stricter choice for a deployed agent. The utterance-level test is silence-matched, so trailing pause duration carries no label information and the decision has to rest on the speech before the pause.

We report five metrics here. Each endpoint is correct, a false alarm, or a miss, and \textbf{F1} (the \emph{frame-level F1}) is the harmonic mean of the resulting precision and recall. \textbf{FA\%} is the fraction of utterances whose first trigger is premature; \textbf{Miss\%} the fraction with no detection; \textbf{FA/utt} the mean premature triggers per utterance; \textbf{Med.\ Lat.} the median $\delta$ over correct detections. 
% FA\% and FA/utt capture different failure modes: whether the first response was premature versus how often the model misfires.

A second, \emph{utterance-level setting} scores the pooled classifiers: each receives one segment that ends in a pause and decides whether the turn is complete. Silence length alone could separate a true turn end from a mid-turn pause, so we match the two classes on trailing silence. A complete segment ends 150\,ms after the true turn end and an incomplete one at a real mid-turn pause with the same 150\,ms taken from earlier in the turn, so the decision must come from the speech before the pause. We score this setting with AUROC, which is threshold-free and imbalance-robust, so it needs no per-set operating point and cannot be inflated by favoring the majority class.

\section{Results and Discussion}

We now ask whether the genre of the training data changes how a streaming detector finds the end of a turn. Every system is scored at its own best-F1 threshold on each dataset, under the same metrics and tolerance window. The numbers are therefore an upper bound rather than a deployable result, but the same tuning is applied to our model and to every baseline, so the comparison is fair and what it settles is the ranking between systems.

\subsection{The Two Genres Differ Mainly in Within-Turn Events}
\label{sec:data_analysis}

\begin{table}[t]
\caption{Dataset characteristics (pauses $\geq$100\,ms, long pauses $\geq$400\,ms, from the pipeline's VAD segments).}
\label{tab:data_stats}
\centering
\footnotesize
\setlength{\tabcolsep}{3.5pt}
\begin{tabular}{lccccc}
\toprule
 & Conv-10K & ST-10K & Int-5K & CASPER & ST-C \\
\midrule
Mean dur.\ (s)    & 11.85 & 7.51  & 11.79 & 4.99  & 9.12 \\
Median dur.\ (s)  & 10.90 & 7.04  & 10.80 & 3.33  & 9.76 \\
Pauses / utt      & 8.37  & 5.13  & 8.47  & 6.60  & 6.00 \\
Long pauses / utt & 2.10  & 1.22  & 2.12  & 1.34  & 1.45 \\
Silence ratio     & 0.687 & 0.333 & 0.688 & 0.600 & 0.433 \\
\bottomrule
\end{tabular}
\end{table}

Before comparing models, we look at what actually separates the two training genres, because this is what a detector must learn and it predicts where the genres should diverge. Table~\ref{tab:data_stats} reports pause statistics for the training and evaluation sets.

\noindent\textbf{The genres differ in one thing that matters, how many pauses fall inside a turn.} Conv-10K carries far more within-turn structure than ST-10K, with roughly 60\% more internal pauses per utterance, nearly double the silence ratio, and over 70\% more long pauses. These are the ordinary marks of spontaneous speech, the hesitations, breath pauses, and mid-turn silences, and a detector has to learn to sit through them rather than read them as turn ends. Conv-10K utterances run longer too, so each example gives more context to hold across.

\noindent\textbf{The evaluation sets spread along the same axis.} Int-5K matches Conv-10K, CASPER keeps the high silence ratio of spontaneous conversation despite its shorter utterances, and ST-Complete falls between the two training distributions. What changes across genres is not what a pause looks like but how many pauses fall \emph{within} a turn, and that is the one distinction conversational training teaches and single-turn training does not.

\subsection{Conversational Training Generalises Across Domains}
\label{sec:conv_gen}

\begin{table}[t]
  \centering
  \caption{ConvLSTM trained on conversational (Conv-10K) vs.\ single-turn (ST-10K) data, scored frame-level on four test sets under the protocol of Sec.~\ref{sec:metrics}. Best F1 per test set in bold. Med.\ Lat.\ is the median latency over correct detections.}
  \label{tab:data_type}
  \footnotesize
  \setlength{\tabcolsep}{4pt}
  \begin{tabular}{l l c c c c c}
  \toprule
  Test & Train & F1 & Miss\% & FA\% & FA/utt & Med.\ Lat. \\
  \midrule
  \multirow{2}{*}{Int-5K}
    & Conv-10K & \textbf{0.894} & 0.1 & 19.1 & 1.22 & 190ms \\
    & ST-10K   & 0.444 & 0.0 & 71.5 & 5.72 & 20ms \\
  \midrule
  \multirow{2}{*}{SLURP}
    & Conv-10K & \textbf{0.760} & 9.6 & 29.1 & 0.49 & 140ms \\
    & ST-10K   & 0.558 & 20.2 & 41.1 & 1.04 & 20ms \\
  \midrule
  \multirow{2}{*}{CASPER}
    & Conv-10K & \textbf{0.959} & 1.3 & 6.6 & 0.07 & 400ms \\
    & ST-10K   & 0.694 & 2.2 & 44.7 & 1.88 & 260ms \\
  \midrule
  \multirow{2}{*}{ST-Comp}
    & Conv-10K & \textbf{0.911} & 0.9 & 15.4 & 0.64 & 250ms \\
    & ST-10K   & 0.738 & 0.8 & 40.7 & 2.31 & 50ms \\
  \bottomrule
  \end{tabular}
\end{table}

With the genres characterised, the training genre is the only thing we change. We train the same detector on Conv-10K and on ST-10K and test both on all four sets (Table~\ref{tab:data_type}).

\noindent\textbf{Conversational training wins on every test set, in domain and out.} Conv-10K gives higher F1 than ST-10K on all four sets, and the margin holds even on SLURP and ST-Complete, the two sets closest to the single-turn data. The advantage is therefore not a training-test domain match; it travels with the training genre. Where, then, does the single-turn model fall behind?

\noindent\textbf{The gap comes from false alarms, not misses.} The single-turn model finds the turn end in almost every utterance, as its low miss rate shows, but triggers too early. Trained on complete single-turn utterances with little exposure to mid-turn pauses, it reads the hesitations and silences of natural speech as turn ends. This is clearest on CASPER, the most spontaneous set, where the conversational model holds false alarms to 6.6\% against 44.7\% for the single-turn model. ST-10K reaches lower median latency only by firing at the first pause, leaving it two to five times as many premature triggers per utterance. In a live system a premature trigger interrupts the speaker, the costlier error, so cutting false alarms is what matters most in practice.

\noindent\textbf{The benefit follows the genre, not the domain.} CASPER also rules out a domain-similarity explanation. It is an independent open-source corpus recorded under different conditions from our call-centre data, so it is out of domain for both training sets. The conversational model still wins there by a wide margin, so the gain comes from the conversational genre and not from any resemblance to our own recordings, just as the pause statistics in Section~\ref{sec:data_analysis} would predict.

\noindent\textbf{The gaps are large and statistically reliable.} A paired McNemar test~\cite{Dietterich1998Approximate} over the utterances where the models disagree gives $p<10^{-70}$ on all four sets, with the conversational model right far more often, and a 5{,}000-sample bootstrap~\cite{Efron1993Bootstrap} puts every 95\% CI on the F1 gap above zero ($+0.17$ to $+0.45$). Nor is this an effect of per-model tuning: a shared threshold $\tau{=}0.7$ leaves every gap at least as large ($+0.49$, $+0.31$, $+0.32$, $+0.20$ on Int-5K, SLURP, CASPER, ST-Complete).

\subsection{At the Utterance Level, the Benefit Holds Only on Conversational Speech}
\label{sec:data_vs_arch}

\begin{figure}[!t]
  \centering
  \includegraphics[width=\columnwidth]{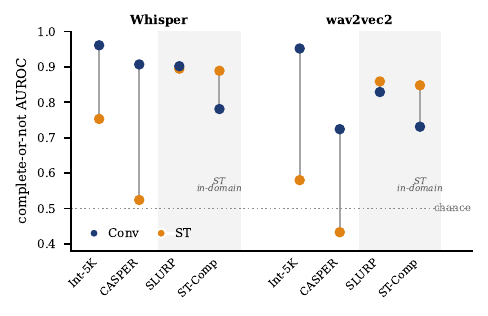}
  \caption{Utterance-level complete-or-not AUROC for two frozen encoders, conversational (Conv) vs single-turn (ST) training, across four test sets. Dotted line: chance.}
  \label{fig:disentangle}
  \vspace{-0.4cm}
\end{figure}

We established the benefit of conversational training for a frame-level detector. Most recent systems decide at the utterance level instead, classifying a segment that ends in a pause as complete or not, so the question is whether the benefit survives there. We repeat the comparison in that setting, scoring the two pooled classifiers by AUROC on the matched segments of Section~\ref{sec:metrics} (Figure~\ref{fig:disentangle}).

\noindent\textbf{On conversational speech, the benefit holds for both encoders.} Conversational training raises AUROC substantially on both conversational sets, by $+0.21$ (Int-5K) and $+0.38$ (CASPER) for Whisper-Tiny and by $+0.37$ and $+0.29$ for wav2vec2-base (Figure~\ref{fig:disentangle}). On CASPER the single-turn head is near or below chance (AUROC 0.52 and 0.43), so conversational training accounts for almost all of the separation. The two encoders share no pretraining, and CASPER is out of domain for both, so the gain reflects genre rather than domain.

\noindent\textbf{On single-turn speech, the benefit is neutral or domain-bound.} On SLURP, out of domain for both training sets, the two heads are effectively tied ($\Delta$AUROC $+0.01$ and $-0.03$), so conversational training does not cost single-turn ability. On ST-Complete the single-turn head wins, but that set shares its source with the single-turn training data (SmartTurn v3), so its head is in domain while the conversational head is not; even there the conversational head stays well above chance (AUROC $0.78$ and $0.73$). Both single-turn sets conclude: the ST-Complete gap reflects domain match, not genre effect.

\noindent \textbf{The gaps are significant.} A paired bootstrap over utterances places a 95\% confidence interval on each AUROC gap. The four conversational gains are clear of zero, as are the two ST-Complete reversals, while the SLURP gaps sit at zero. So conversational training gives a real and large utterance-level advantage on conversational speech, stays neutral on out-of-domain single-turn speech, and loses only where the single-turn model is in domain.

\begin{table*}[!t]
  \caption{Cross-dataset results at each model's best threshold. Bold = highest F1. Latency in ms; parentheses give the utterance count; FA/utt (Section~\ref{sec:metrics}) is omitted for space.}
  \label{tab:crossdataset}
  \centering
  \footnotesize
  \setlength{\tabcolsep}{3.5pt}
  \begin{tabular}{l cccc cccc cccc}
  \toprule
  & \multicolumn{4}{c}{Int-5K}
  & \multicolumn{4}{c}{CASPER}
  & \multicolumn{4}{c}{ST-Complete} \\
  \cmidrule(lr){2-5}\cmidrule(lr){6-9}\cmidrule(lr){10-13}
  Model
  & F1 & FA\% & Miss\% & Med.\ Lat.
  & F1 & FA\% & Miss\% & Med.\ Lat.
  & F1 & FA\% & Miss\% & Med.\ Lat. \\
  \midrule
  ConvLSTM (ours)
  & \textbf{.894} & 19.1 &  0.1 & 190\,(4040)
  & \textbf{.959} &  6.6 &  1.3 & 400\,(2301)
  & \textbf{.911} & 15.4 &  0.9 & 250\,(2092) \\
  \midrule
  Silero VAD
  & .781 & 33.2 &  2.7 &  50\,(3205)
  & .734 & 29.6 & 12.4 &  12\,(1448)
  & .673 & 13.2 & 36.1 & 110\,(1268) \\
  WebRTC agg3
  & .476 & 66.7 &  2.1 &  20\,(1560)
  & .512 & 52.8 & 12.8 &   0\,(861)
  & .420 & 41.9 & 31.5 &  90\,(665) \\
  \midrule
  SmartTurn v3
  & .523 & 28.9 & 35.7 &  48\,(1770)
  & .390 & 22.0 & 53.7 &   8\,(606)
  & .652 & 50.6 & 12.3 & 150\,(928) \\
  SmartTurn FT
  & .525 & 64.0 &  0.4 &  50\,(1780)
  & .644 & 44.6 &  7.9 &   4\,(1188)
  & .694 & 34.6 &  1.0 & 152\,(1610) \\
  \bottomrule
  \end{tabular}
\end{table*}

\begin{figure}[t]
\centering
\includegraphics[width=\columnwidth]{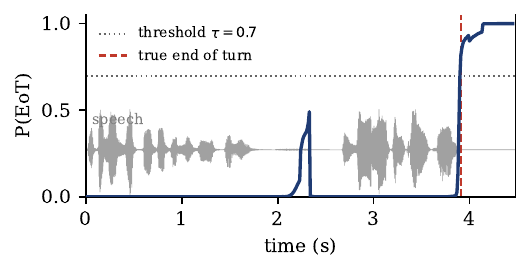}
\caption{Frame-level P(EoT) on a CASPER turn, with the speech waveform overlaid and the true end dashed.}
\label{fig:qualitative}
\end{figure}

\noindent \textbf{Why the test genre matters.} This is why the effect has gone unnoticed. Utterance-level classifiers are usually benchmarked on single-turn or command data~\cite{smartturn2025,bastianelli2020}, exactly the regime where the conversational benefit does not surface. It is not that conversational training fails there. A single decision on a clean ending simply does not exercise the within-turn events the two genres differ on. This is consistent with the frame-level result, where the conversational model wins even on the single-turn test sets, because that metric scores the whole streaming recording and rewards holding back through every mid-turn pause (Figure~\ref{fig:qualitative}), whereas the utterance-level decision sees only one fixed ending. The practical lesson is that the utterance-level benchmark itself decides whether the training-genre effect is visible, so evaluating only on single-turn or command speech leaves it untested.

\subsection{Outperforming Open-Source Baselines with Fewer Parameters}
\label{sec:baselines_results}

\noindent\textbf{The ConvLSTM leads on all three sets at a fraction of SmartTurn's size.} With the genre effect established at both granularities, we benchmark it against open-source systems on the three test sets that carry conversational or multi-turn structure (Table~\ref{tab:crossdataset}). SLURP, a single-command set, appears only for our own model, in Table~\ref{tab:data_type}. It leads in F1 on all three at 1.8M parameters, about 4.4$\times$ fewer than the 8M-parameter SmartTurn, and it is the only system to hold false alarms, misses, and latency in check at once.

\noindent\textbf{The two silence VADs fail exactly as their design predicts.} On CASPER the flowing dialogue leaves few clean silence gaps, so they miss most turns, while on ST-Complete the frequent mid-speech pauses drive their false alarms up. They reach low latency only by firing at the first silence, so their speed comes at the cost of reliability. The ConvLSTM instead waits about 400\,ms on CASPER, the price of holding through mid-turn pauses rather than stopping at them, which sits near the upper end of natural between-turn gaps~\cite{levinson2015,heldner2010} while keeping both false alarms and misses low.

\noindent\textbf{Reading the speech does not by itself beat the silence rule.} Trained on single-turn data, SmartTurn~v3 transfers poorly to these conversational endings and trails even Silero on all three sets, missing most turns (Miss\% up to 53.7 on CASPER), while fine-tuning only trades the misses for over-firing (FA\% up to 64 on Int-5K). A learned endpoint detector thus rises or falls with its training genre, the factor we isolate throughout.

\subsection{Limitations}
\label{sec:limitations}
A few limitations frame our conclusions. The conversational corpus is English-only telephony while the single-turn data is multilingual and partly synthetic~\cite{smartturnv3blog}, so the advantage may not hold across languages, accents, or styles, and part of the gap could reflect synthetic-speech artifacts rather than genre. Labels for SLURP and ST-Complete are VAD-derived rather than human-annotated, which may favor some model types. The frozen Whisper and wav2vec2 comparators receive 8\,kHz audio upsampled to 16\,kHz, lowering their absolute scores, though this applies to both genres and leaves the Conv-minus-ST comparison intact. The two training sets differ in genre and source domain, so part of the gap could be domain rather than genre, though the out-of-domain CASPER result argues against this. Our internal sets are private, but the effect reproduces on public data, since CASPER, the single-turn data, and both frozen comparators are all public. Finally, we study the two genres for which abundant real speech can be reliably labelled today; extending to others such as multi-turn human-machine dialogue awaits suitable corpora~\cite{mmf2f2025}.

\section{Conclusion}
This paper treats speech genre as a primary experimental variable for end-of-turn detection. Across four test sets, a streaming ConvLSTM trained on conversational human-human data steadily beats the same model trained on single-turn human-machine data, with the gains coming from fewer false alarms on the mid-turn pauses and hesitations that single-turn corpora rarely contain. The benefit holds at both decision granularities: a frame-level model captures it directly, and an utterance-level classifier captures it too, but only on conversational test speech. On single-turn benchmarks the two genres tie out of domain and the single-turn model leads only on its own source domain, so such evaluations leave the effect untested rather than contradict it. Against open-source baselines, the proposed model attains the highest F1 on all test sets with far fewer parameters than the learned SmartTurn baseline. These results suggest that who the speaker is talking to shapes the training signal as much as model capacity or feature choice, and that speech genre deserves attention as a first-class design decision for EoT detection. Natural next steps are multilingual settings and mixtures of conversational and single-turn data.

% \section{Acknowledgments}

\bibliographystyle{IEEEtran}
\bibliography{mybib}

\end{document}